\documentclass[10pt,conference]{IEEEtran}

\usepackage{xcolor}

\usepackage{multirow} 

\usepackage[pdftex]{graphicx}

\IEEEoverridecommandlockouts
\makeatletter
\def\ps@IEEEtitlepagestyle{
  	\def\@oddfoot{\Footer} 
	}
\let\old@ps@headings\ps@headings
\let\old@ps@IEEEtitlepagestyle\ps@IEEEtitlepagestyle
\def\confheader#1{%
  \def\ps@headings{%
    \old@ps@headings%
    \def\@oddhead{\strut\hfill#1\hfill\strut}%
    \def\@evenhead{\strut\hfill#1\hfill\strut}%
    \def\@oddfoot{\Footer} 
  }%
  \def\ps@IEEEtitlepagestyle{%
    \old@ps@IEEEtitlepagestyle%
    \def\@oddhead{\strut\hfill#1\hfill\strut}%
    \def\@evenhead{\strut\hfill#1\hfill\strut}%
  }%
  \ps@headings%
}
\makeatother
\def\Footer{
    {\footnotesize
   \begin{minipage}{\textwidth}
   \centering \scriptsize{DISTRIBUTION STATEMENT A. Approved for public release: distribution is unlimited.\\
   }  \end{minipage}
  }
 }
\newcommand{\subf}[2]{%
  {\begin{tabular}[t]{@{}c@{}}
  #1\\#2
  \end{tabular}}%
}

\begin{document}

\title{Cross-Scale Persistence Analysis of EM Side-Channels for Reference-Free Detection of Always-On Hardware Trojans\\
}
\author{\IEEEauthorblockN{Mahsa Tahghigh}
\IEEEauthorblockA{Electrical Engineering and Computer Science\\
Howard University,\\
Washington DC, USA,\\
mahsa.tahghigh@bison.howard.edu}
\and
\IEEEauthorblockN{Hassan Salmani, Ph.D.}
\IEEEauthorblockA{Electrical Engineering and Computer Science\\
Howard University,\\
Washington DC, USA,\\
hassan.salmani@howard.edu}
}
\maketitle

\begin{abstract}
Always-on hardware Trojans pose a serious challenge to integrated circuit trust, as they remain active during normal operation and are difficult to detect in post-deployment settings without trusted golden references. This paper presents a reference-free detection framework based on cross-scale persistence analysis of electromagnetic (EM) side-channels, targeting always-on parasitic hardware behavior. The proposed method analyzes EM emissions across multiple time–frequency resolutions and constructs stability maps that capture the consistency of spectral features over repeated executions. Gaussian Mixture Models (GMMs) with Bayesian Information Criterion (BIC)–based model selection are used to characterize statistical structure at each scale. We introduce cross-scale saturation, variability, and median mixture complexity metrics that quantify whether statistical structure evolves naturally or remains persistently anchored across resolutions. Experimental results on AES implementations show that Trojan-free designs exhibit scale-dependent variability consistent with transient switching behavior, while always-on Trojans produce persistent statistical signatures that suppress cross-scale evolution. Furthermore, different Trojan classes - such as workload-correlated leakage-information Trojans and independent ring-oscillator Trojans - exhibit distinct persistence patterns. These findings demonstrate that cross-scale persistence provides a physically interpretable and robust assurance signal for unsupervised, reference-free detection of always-on hardware Trojans.
\\
\end{abstract}
\renewcommand\IEEEkeywordsname{Keywords}

\section{Introduction}
Ensuring the trustworthiness of integrated circuits has become a critical challenge as hardware systems are increasingly deployed in security- and mission-critical environments \cite{DoDI_5200_50}\cite{PrabhakaraRao2025}. Among the most difficult threats to detect are always-on hardware Trojans (HTs), which do not rely on rare trigger conditions and may remain active throughout normal operation. Unlike trigger-based HTs \cite{tahghigh2024gmm}\cite{10.1145/3802543}\cite{TahghighArXiv2026b}\cite{11014346}, always-on HTs continuously influence circuit behavior, potentially leaking sensitive information or degrading system integrity while avoiding obvious functional failures \cite{TahghighArXiv2026}. Their persistent nature makes them especially problematic in post-deployment scenarios, where invasive inspection, exhaustive testing, or access to trusted golden references is often impractical\cite{Bhunia2018HardwareTrojan}\cite{Salmani2018}. 

Electromagnetic (EM) side-channel analysis has emerged as a powerful non-invasive technique for observing internal circuit activity \cite{10752788}. He et al. \cite{7994702} proposes a golden-chip-free EM HT detection method that compares measured spectra against a simulation-derived reference generated from trusted RTL. While eliminating the need for a physical golden device, the approach still depends on accurate modeling and trusted design visibility, and its evaluation assumes preset HT activation, limiting applicability to fielded systems. John et al. \cite{11195049} propose a post-deployment HT detector using supervised machine learning on raw power side-channel traces, achieving high accuracy for known HTs. However, the approach depends on labeled HT-infected and HT-free data and may not generalize to unseen HTs or operational drift. Sun et al. \cite{9534884} employ CWT-based EM time–frequency representations combined with transfer learning to classify HT-infected and HT-free designs with high accuracy. Recent work has explored unsupervised learning of electromagnetic (EM) side-channels for hardware Trojan detection, including Deep SVDD–based anomaly detection without labeled Trojan data or golden reference chips \cite{10.1109/TVLSI.2024.3458892}. While effective when Trojan-induced EM signatures are strong, such approaches primarily operate on static spectral features and do not explicitly model the stability or persistence of EM behavior across executions or analysis scales. This limits robustness for always-on or leakage-oriented Trojans, whose effects may manifest as subtle perturbations rather than distinct spectral peaks.

\begin{figure*}[t]
    \centering
    \includegraphics[width=\linewidth]{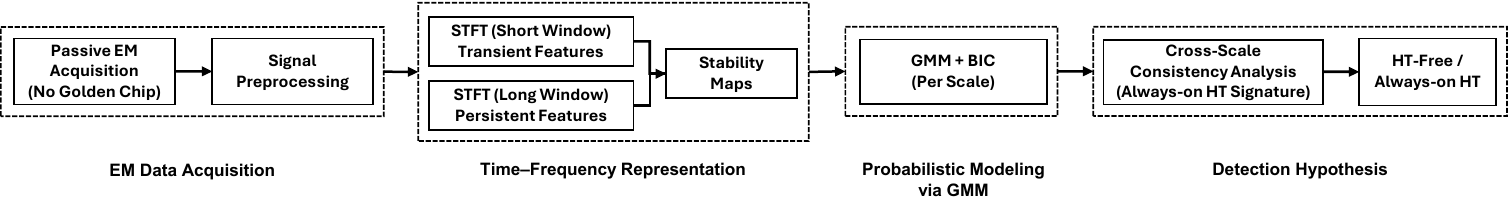}
    \caption{Overview of the proposed reference-free EM side-channel analysis framework for detecting always-on HTs. EM traces are collected during repeated executions and transformed into multi-scale time–frequency representations. Stability maps are constructed to capture persistent spectral structure across executions and window sizes. Gaussian Mixture Models (GMMs) with bounded model order are used to characterize the statistical structure at each scale. Detection is based on cross-scale persistence analysis, where always-on HTs produce persistent or saturated mixture structures with reduced variability across resolutions, in contrast to HT-free designs that exhibit adaptive, scale-dependent statistical behavior.}
    \label{fig:flow}
\end{figure*}

More importantly, many existing approaches implicitly assume that malicious behavior manifests as distinct spectral signatures at a single resolution, whereas always-on HTs primarily alter how statistical structure persists across scales, rather than introducing isolated features. Prior circuit-level studies have shown that post-fabrication hardware behavior may diverge from design-time assumptions, complicating validation in deployed systems \cite{cta3980}. These challenges motivate the need for reference-free and unsupervised detection methods that operate under realistic post-deployment conditions and rely solely on observed physical behavior rather than trusted models or prior knowledge of HT implementations. 

Recent years have seen growing interest in applying machine learning (ML) and artificial intelligence (AI) techniques to hardware Trojan detection, particularly using side-channel information such as power and electromagnetic (EM) emissions. Both supervised and unsupervised models—including support vector machines, deep autoencoders, and one-class anomaly detectors—have demonstrated promising detection accuracy under controlled experimental settings. However, many ML-based approaches implicitly rely on strong feature separability, carefully curated training data, or extensive hyperparameter tuning, and often lack physical interpret ability of the learned decision boundaries. Moreover, their generalization to unseen Trojan implementations, varying workloads, or post-deployment operating conditions remains an open challenge \cite{10.1109/TVLSI.2024.3458892}\cite{10.1007/s10836-018-5726-9}.

To address these limitations, in this paper, we propose a reference-free framework based on cross-scale persistence analysis of EM side-channels for detecting always-on HTs. The approach analyzes EM emissions across multiple time–frequency resolutions and constructs stability maps that capture the consistency of spectral features over repeated executions. Gaussian Mixture Models (GMMs) with Bayesian Information Criterion (BIC)–based model selection are used to characterize the statistical structure of these stability maps at each scale \cite{McLachlanPeel2000}\cite{10.1214/aos/1176344136}. 
Rather than relying solely on average mixture complexity, we introduce cross-scale saturation and variability metrics that quantify whether statistical structure evolves naturally or remains persistently anchored across resolutions.

We evaluate the proposed framework on AES implementations under three conditions: HT-free operation, AES augmented with a workload-correlated leakage-information HT, and AES augmented with an independent ring-oscillator HT. The results show that HT-free designs exhibit adaptive, scale-dependent statistical behavior, while always-on HTs suppress this evolution in distinct ways depending on their interaction with the workload. These findings demonstrate that cross-scale persistence serves as a robust and interpretable assurance signal, enabling unsupervised detection of always-on HTs without golden references or labeled data.

\section{Technical Approach}
We consider a threat model in which an adversary inserts an always-on HT into a digital integrated circuit during design, fabrication, or third-party IP integration. Unlike trigger-based HTs that activate under rare conditions, an always-on HT continuously introduces parasitic activity—such as additional switching, leakage paths, or covert modulation—during normal operation. This persistent behavior enables long-term information leakage or system degradation while avoiding detectable functional failures, making always-on HTs particularly challenging to identify through conventional logic testing or runtime checks. 

We assume a post-deployment detection scenario in which the device under test is already fabricated and operational, and no trusted golden reference chip, labeled HT-infected dataset, or accurate simulation-derived EM model is available. The proposed framework aims to detect always-on HTs by analyzing the persistence of statistical structure in electromagnetic (EM) side-channels across multiple time–frequency resolutions. The key principle is that benign circuit activity exhibits scale-dependent evolution, while always-on parasitic behavior introduces persistent statistical signatures that resist temporal averaging. To capture this distinction in a reference-free and unsupervised manner, the framework consists of four main steps, as illustrated in Figure \ref{fig:flow}.

\begin{figure}[b]
    \centering
    \includegraphics{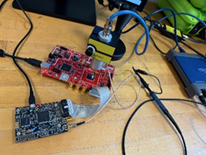}
    \caption{Experimental setup.}
    \label{fig:SetUp}
\end{figure}

\begin{figure*}[t]
    \centering
    \includegraphics[width=\linewidth]{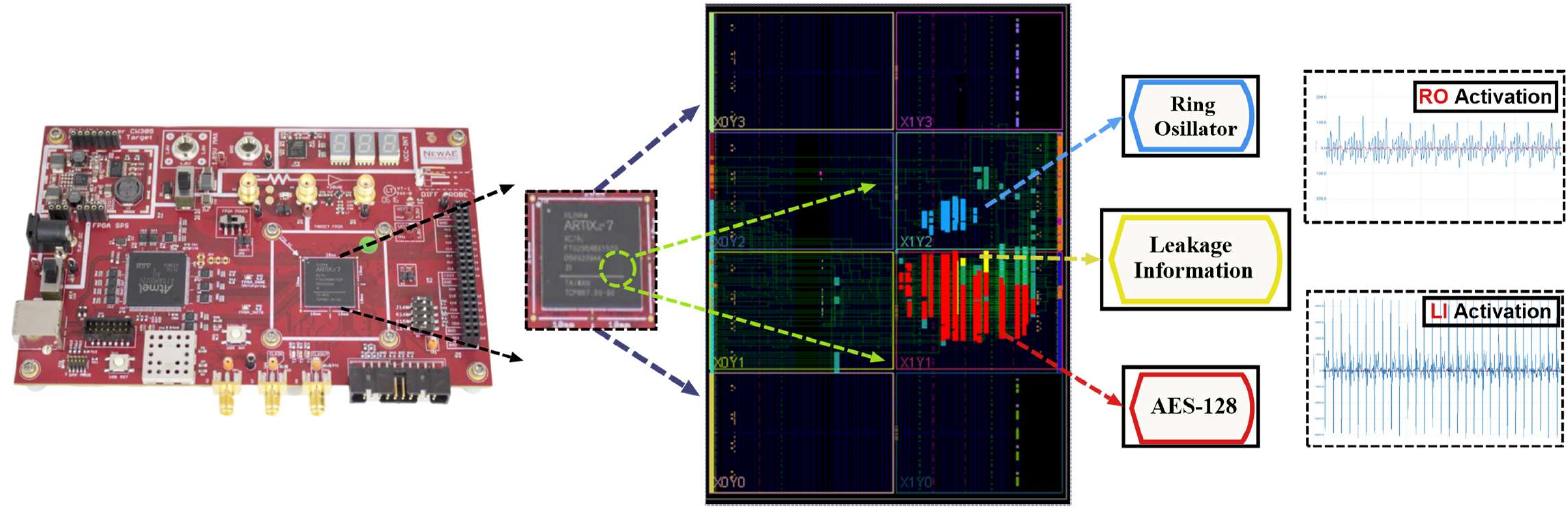}
    \caption{FPGA implementation of the AES-128 core with two always-on hardware Trojans: a workload-correlated leakage-information Trojan (LI-HT) and an independent ring-oscillator Trojan (RO-HT). The figure illustrates circuit placement and representative EM signal activity when each Trojan is active.}
    \label{fig:FPGADesign}
\end{figure*}

\begin{itemize}
    \item \textbf{\textit{EM Data Acquisition}}: We acquire electromagnetic (EM) emanations from a target integrated circuit during normal operation. The process does not assume knowledge of the chip internals or access to a “golden” reference device. Instead, EM signals are collected passively in situ while the device executes its intended functionality, ensuring realistic side-channel conditions where the effects of always-on HTs may be subtle and masked by legitimate circuit activity. 
    \item \textbf{\textit{Time–Frequency Representation and Stability Map Construction}}: To analyze EM behavior across different temporal resolutions, each trace is transformed into a time–frequency representation using the Short-Time Fourier Transform (STFT) \cite{OppenheimSchafer2022}. Multiple window sizes are employed, corresponding to different numbers of clock cycles per window, enabling analysis at varying time–frequency scales. For each window size, the resulting spectrograms from repeated executions are aggregated to compute a stability map, defined as the ratio of the mean spectral magnitude to its variance across executions. The stability map highlights frequency components that exhibit consistent behavior over time, suppressing transient or noisy features while emphasizing persistent spectral structure. Importantly, by varying the STFT window size, the framework explicitly probes how this structure evolves—or fails to evolve—across scales.
    \item \textbf{\textit{Probabilistic Modeling via GMM}}: At each time–frequency scale, feature vectors are constructed from the stability map by pairing frequency values with their corresponding stability magnitudes. These feature vectors are then modeled using Gaussian Mixture Models (GMMs) with full covariance matrices to capture correlated structure in the EM features. The number of mixture components is selected using the Bayesian Information Criterion (BIC), which balances model fit against complexity. To preserve physical interpretability and avoid over-segmentation of inherently non-Gaussian EM data, the maximum allowable number of mixture components is bounded ($k_{max}=10$), which reflects the expectation that only a limited number of dominant emission regimes or parasitic sources can be physically active within a given time–frequency window. Multiple random initializations are used to ensure stability of the selected model order.
    \item \textbf{\textit{Cross-Scale Detection Hypothesis and Metrics}}: Rather than relying solely on the absolute number of mixture components selected at a given scale, detection is based on cross-scale persistence analysis. For each window size, the selected GMM order is recorded across multiple executions, and three complementary metrics are derived:
    \begin{itemize}
        \item \textit{Median mixture complexity}, which provides a coarse summary of statistical structure at each scale.
        \item \textit{Saturation ratio}, defined as the fraction of executions in which the BIC-selected model reaches the maximum allowed number of mixture components, indicating persistent demand for high statistical complexity.
        \item \textit{Within-window variance}, which quantifies the variability of selected model order across executions and reflects the adaptability of the underlying statistical structure.
    \end{itemize}

    The detection hypothesis is that HT-free designs will exhibit scale-dependent variability, characterized by low saturation and higher variance as different operational phases dominate at different resolutions. In contrast, always-on HTs are expected to suppress this natural evolution, producing persistent or anchored statistical structure across scales. Different classes of always-on HTs may manifest this persistence in distinct ways, but all deviate from the adaptive behavior observed in benign designs. 
\end{itemize}

\begin{figure*}[t]
    \centering
    \includegraphics[width=\linewidth]{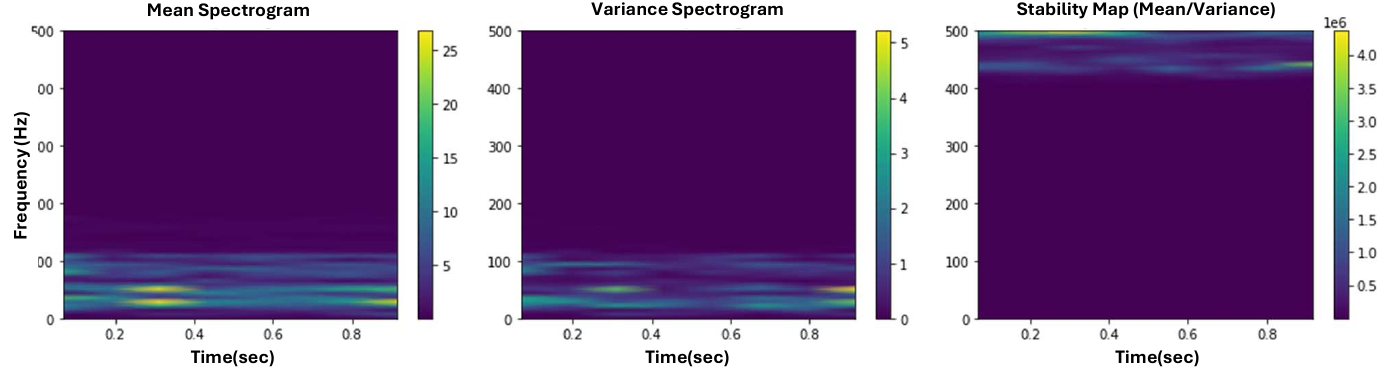}
    \caption{Spectrogram analysis of HT-free AES-128 after applying a fixed key and 500 plaintexts with the segment length of 138 for STFT.}
    \label{fig:SpectrogramHT-FreeAES128-WS138STFT}
\end{figure*}

\begin{figure*}[t]
    \centering
    \includegraphics[width=\linewidth]{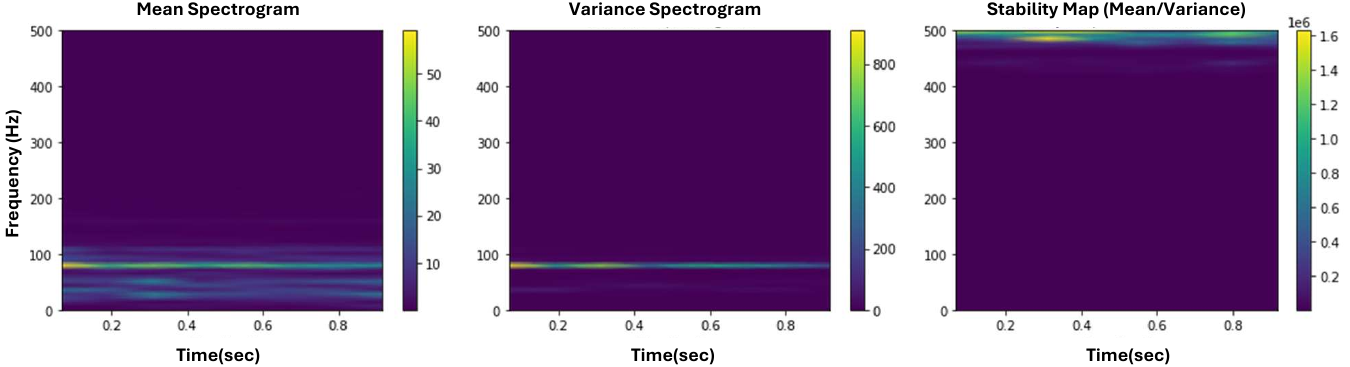}
    \caption{Spectrogram analysis of RO-inserted AES-128 after applying a fixed key and 500 plaintexts with the segment length of 138 for STFT.}
    \label{fig:SpectrogramHT-ROAES128-WS138STFT}
\end{figure*}

By combining multi-scale time–frequency analysis with probabilistic modeling and cross-scale persistence metrics, the proposed approach provides a reference-free and physically interpretable framework for detecting always-on HTs. The methodology does not assume prior knowledge of HT structure, does not require labeled data, and is well suited for post-deployment assurance scenarios where trusted baselines are unavailable. 

\section{Results and Discussion}
To evaluate the proposed reference-free detection framework, we applied it to an AES-128 encryption engine as a representative complex digital workload. The circuit was executed repeatedly under fixed key and plaintext conditions. Figure \ref{fig:SetUp} shows our setup, which includes a Riscure CW305 board and a Riscure HP EM Probe (1.5 mm) \cite{KeysightInspectorSC4}. EM waveforms are captured using PicoScope software and saved in CSV format. 

The AES-128 is tampered with by two HTs inspired by Trust-Hub \cite{6657085}\cite{Shakya2017}. Inspired by Code-Division Multiple Access (CDMA), an leakage-information HT (LI-HT) is based on AES-T1100. It disperses key-bit leakage across multiple clock cycles using a PRNG seeded with the plaintext to generate a CDMA code, which XOR-modulates the key bits. The modulated sequence is sent to a leakage circuit composed of eight flip-flops, emulating a large capacitance to create a covert power side-channel. The other HT is a ring-oscillator (RO) made of 101 NAND gates. The RO-HT runs independently of the AES-128. Figure \ref{fig:FPGADesign} presents the placement of AES, the leakage information, and the ring oscillator circuits inside the Artix-7 FPGA. The figure further shows an example of the EM signal when the RO-HT and the LI-HT are active. As examples of Step 2 of our proposed methodology, Time– Frequency Representation, Figure \ref{fig:SpectrogramHT-FreeAES128-WS138STFT} and Figure \ref{fig:SpectrogramHT-ROAES128-WS138STFT} show the mean and variance spectrum and stability map (segment length = 138) for STFT, comparing HT-free and RO-inserted AES, respectively.

\begin{figure*}[t]
\centering
\begin{tabular}{ccc}
\\
\subf{\includegraphics[width=50mm]{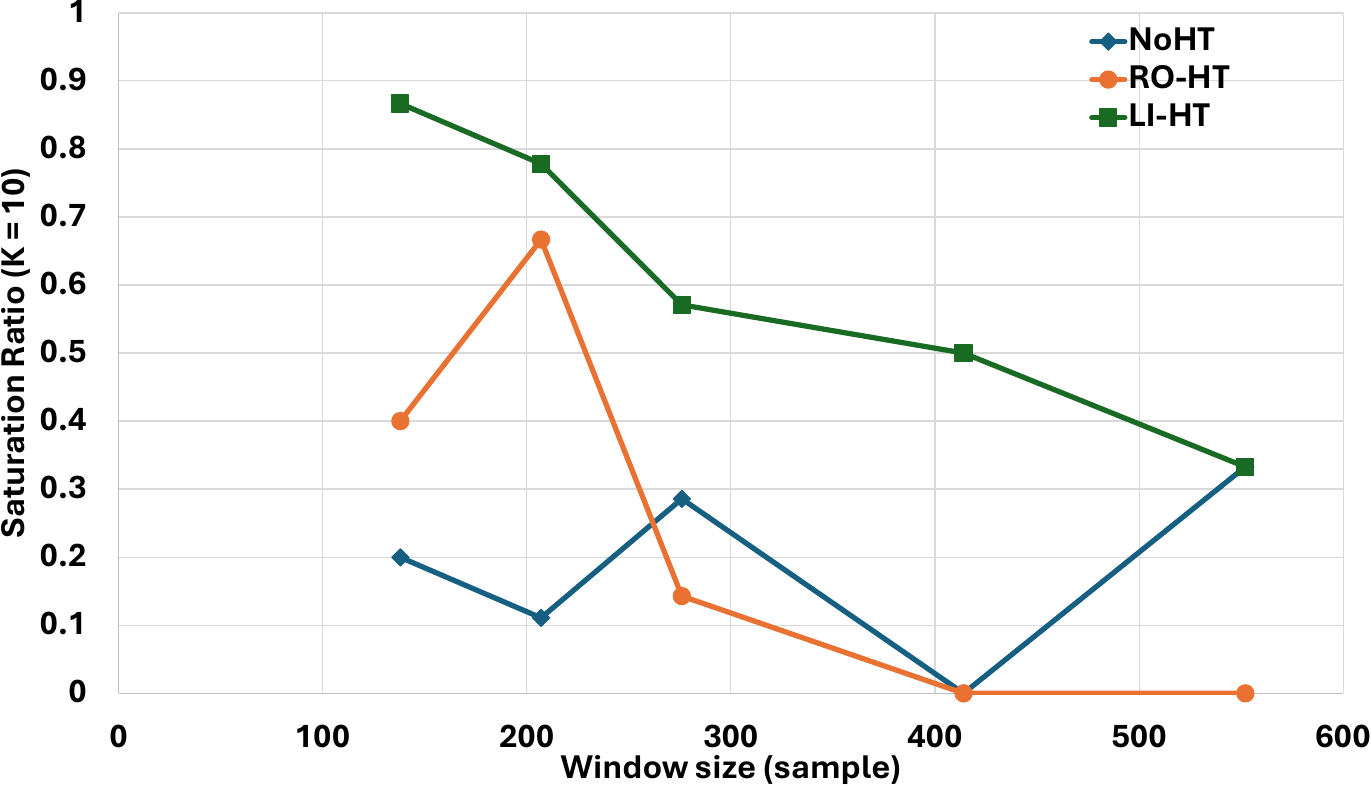}}
{(1)}
&
\subf{\includegraphics[width=50mm]{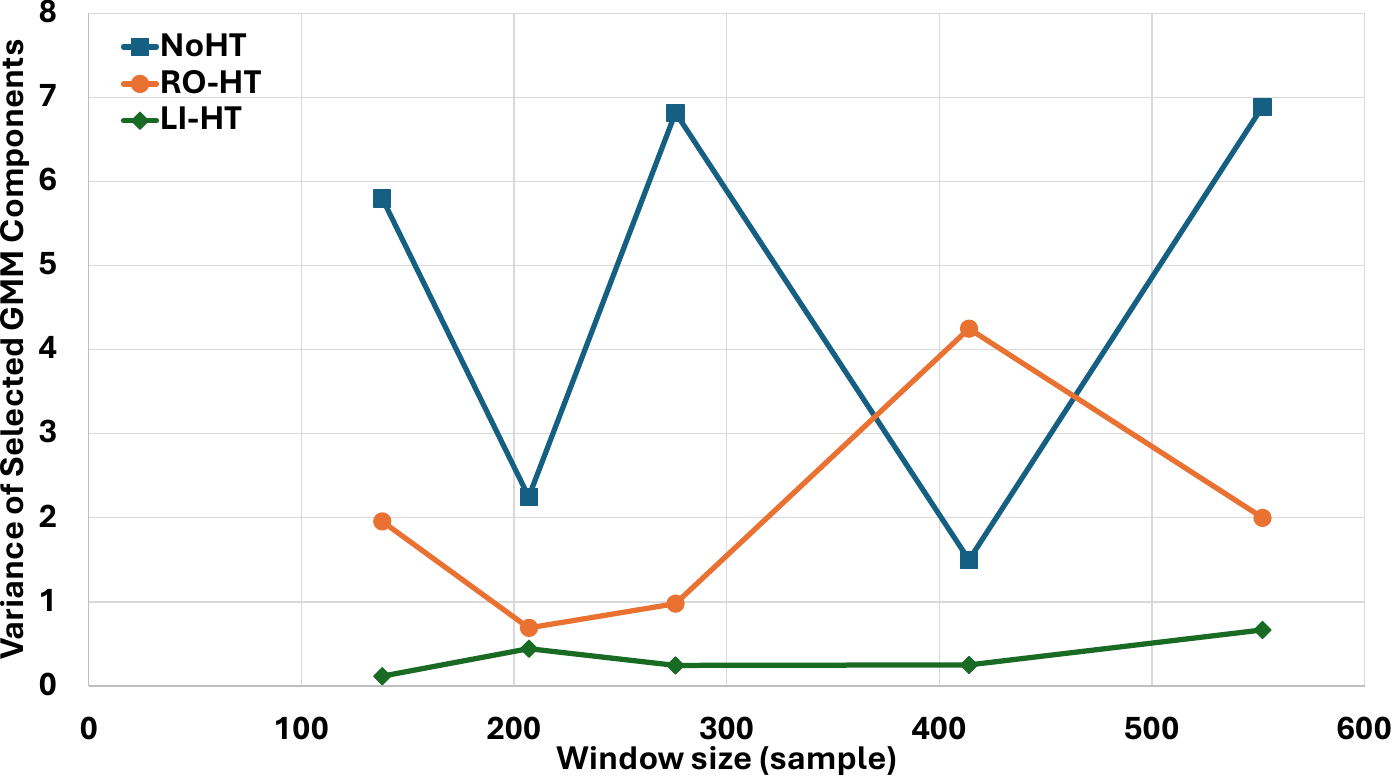}}
{(2)}
&
\subf{\includegraphics[width=50mm]{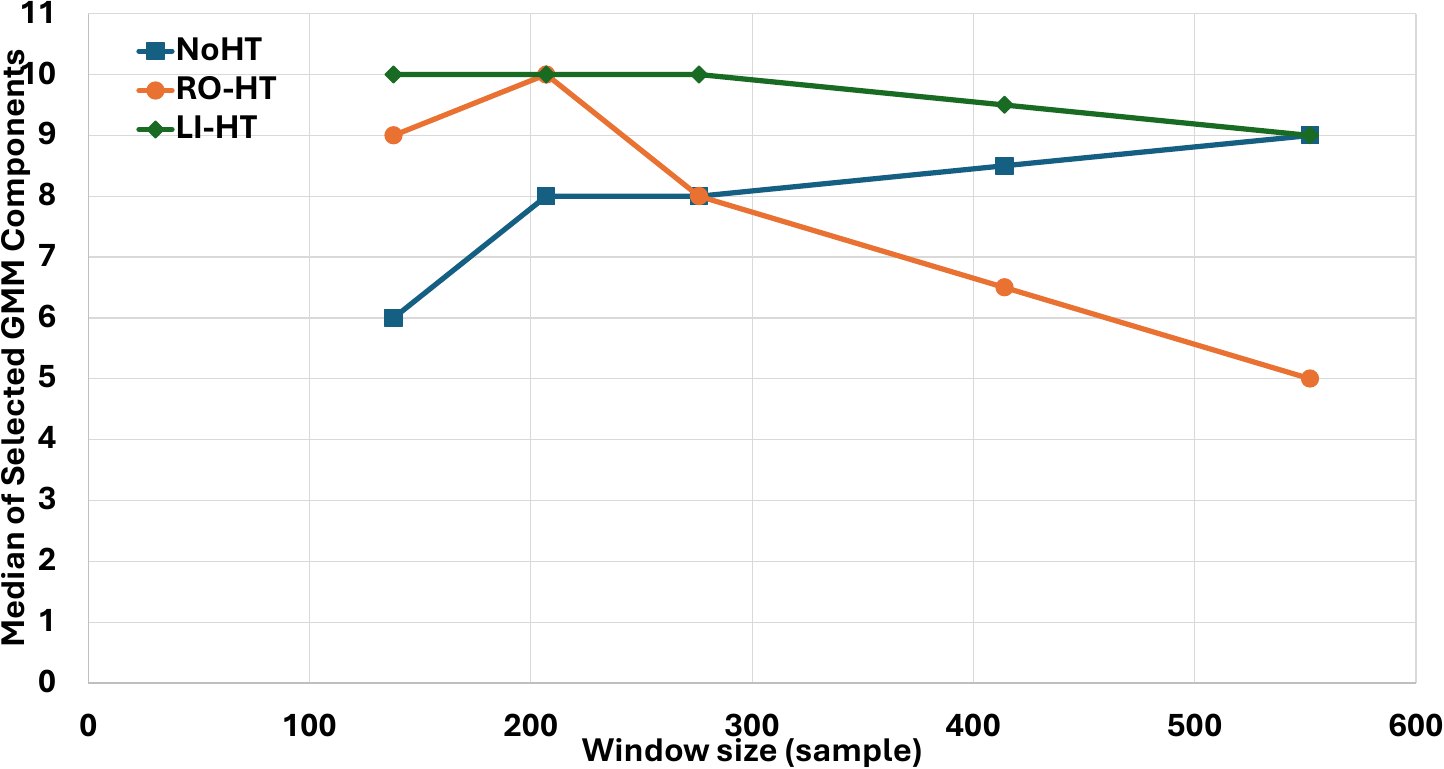}}
{(3)}
\\
\end{tabular}
\caption {Cross-scale behavior of Gaussian Mixture Model (GMM) complexity for HT-free AES (NoHT), AES with a ring-oscillator HT (RO-HT), and AES with a leakage-information HT (LI-HT) across increasing STFT window sizes. (1) Saturation ratio, defined as the fraction of executions in which the BIC-selected model reaches the maximum allowed number of components ($K_{max}=10$); (2) within-window variance of the selected model order across executions; and (3) median number of mixture components. HT-free designs exhibit low saturation and higher variability, reflecting adaptive, scale-dependent statistical structure. In contrast, always-on HTs produce persistent or saturated mixture structures with reduced cross-scale evolution, with distinct trends observed for workload-correlated and independent HT classes.} 
\label {fig:CrossScalePersistenceMetrics}
\vspace{-0.2in}
\end{figure*}

Figure \ref{fig:CrossScalePersistenceMetrics} summarizes the cross-scale behavior of the proposed detection metrics for three configurations: HT-free AES (NoHT), AES with a RO-HT, and AES with a LI-HT. The figure presents three complementary views of mixture complexity across increasing STFT window sizes: (1) saturation ratio, (2) within-window variance of the selected GMM order, and (3) median number of GMM components.

\textbf{\textit{Saturation Ratio}}: Figure \ref{fig:CrossScalePersistenceMetrics}(1) shows the saturation ratio, defined as the fraction of executions in which the BIC-selected GMM reaches the maximum allowable number of mixture components ($k_{max}$=10). The HT-free design consistently exhibits low saturation across all window sizes, indicating that its statistical structure adapts naturally as the analysis scale changes. This behavior is consistent with benign AES execution, where transient and phase-dependent switching activity dominates, and no persistent parasitic source enforces a fixed complexity.
In contrast, both HT-inserted designs demonstrate elevated saturation, but with distinct cross-scale trends. The LI-HT exhibits consistently high saturation across all window sizes, reflecting persistent workload-correlated parasitic activity that remains statistically dominant regardless of temporal resolution. The RO-HT shows high saturation at smaller window sizes, followed by a gradual decrease as the window size increases. This decay reflects the stationary and periodic nature of the ring oscillator, whose contribution becomes increasingly averaged out at coarser time scales.

These results confirm that saturation ratio serves as a direct indicator of persistent parasitic emission sources, and that its cross-scale evolution differentiates HT classes without requiring explicit HT signatures.

\textbf{\textit{Within-Window Variance}}: Figure \ref{fig:CrossScalePersistenceMetrics}(2) presents the variance of the selected GMM order across executions for each window size. The HT-free configuration exhibits relatively higher variance, particularly at finer resolutions, indicating that the statistical structure of EM stability maps changes from execution to execution. This variability reflects the natural diversity of transient switching behavior during AES operation.
The LI-HT demonstrates consistently low variance across all window sizes, indicating that the mixture structure is highly stable and repeatable. This suppression of variability is a hallmark of always-on, workload-correlated parasitic behavior, which constrains the statistical structure across executions. The RO-HT exhibits intermediate behavior, with lower variance at small window sizes and increasing variance at larger windows, reflecting competition between stationary oscillatory emissions and AES activity as temporal averaging increases.

Together, these observations show that variance captures adaptability, complementing saturation by revealing whether mixture complexity is free to evolve or remains statistically anchored.

\textbf{\textit{Median Mixture Complexity}}: Figure \ref{fig:CrossScalePersistenceMetrics}(3) shows the median number of GMM components selected at each window size. While the median alone does not always provide strict separation between configurations, it offers useful contextual information when interpreted alongside saturation and variance. The HT-free design exhibits non-monotonic median behavior, reflecting scale-dependent dominance of different AES phases. The LI-HT remains pinned near the maximum model order, reinforcing the interpretation of persistent, correlated parasitic structure. The RO-HT shows a gradual decrease in median complexity with increasing window size, consistent with the diminishing influence of stationary periodic emissions at coarser scales.

\begin{figure}[t]
    \centering
    \includegraphics[width=8cm]{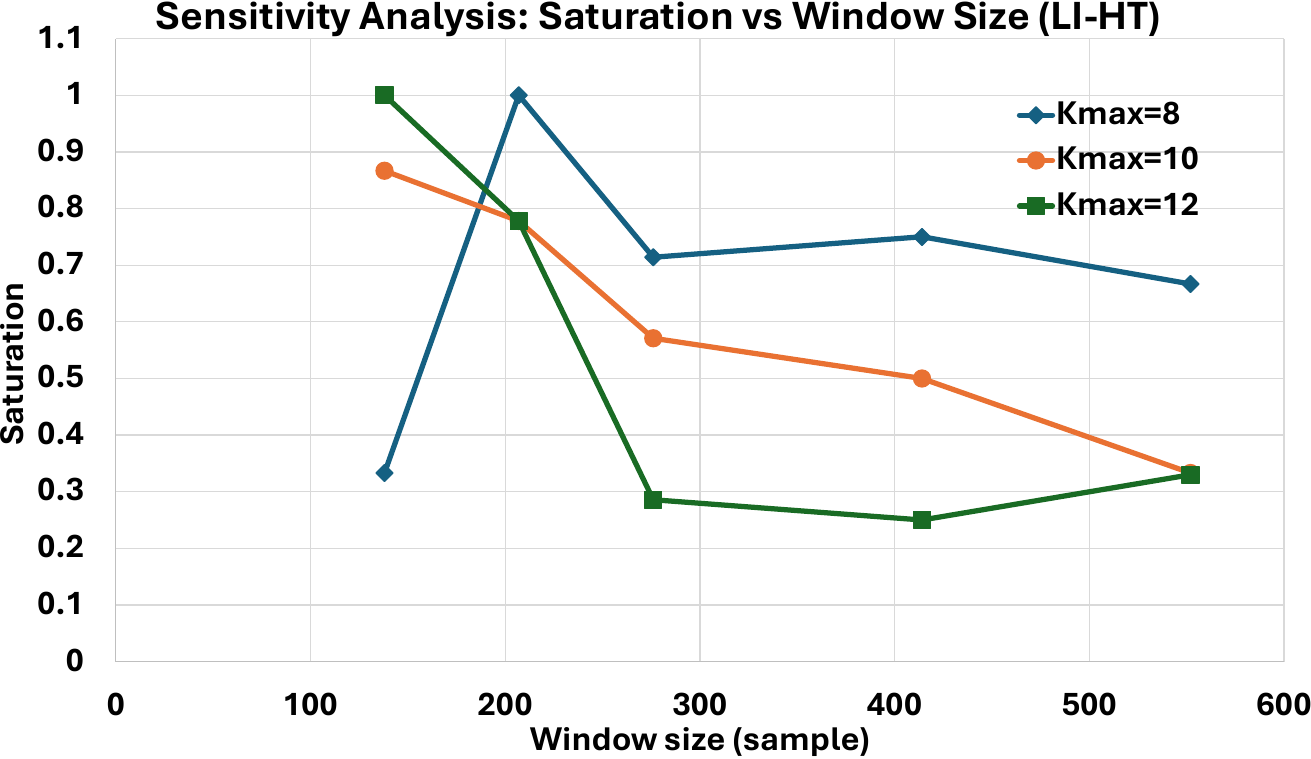}
    \caption{Sensitivity analysis of saturation ratio under varying GMM capacity bounds for for the leakage-information Trojan (LI-HT). Saturation ratio versus STFT window size for the LI-HT evaluated with different maximum numbers of GMM components $k_{max}= 8, 10, 12$.}
    \vspace{-0.1in}
    \label{fig:SensitivityAnalysisSaturationvsWindowSizeLI-HT}
\end{figure}

\textbf{\textit{Sensitivity Analysis}}: We evaluated the robustness of the proposed metrics under varying GMM 
capacity bounds ($K_{max} \in {8,10,12}$). While absolute mixture counts and saturation levels shift with model capacity, the qualitative cross-scale behavior remains consistent across all tested $k_{max}$ values, confirming that detection is driven by persistence trends rather than parameter-specific tuning. HT-free designs continue to exhibit lower saturation and higher variability, LI-HTs remain persistently saturated with low variance, and RO-HTs demonstrate scale-dependent decay. For example, Figure \ref{fig:SensitivityAnalysisSaturationvsWindowSizeLI-HT} presents the sensitivity analysis for saturation versus window size for LI-HT. 

Importantly, these results illustrate that absolute mixture counts are insufficient as a standalone detection metric, but become meaningful when placed in the context of cross-scale persistence and variability. 
The median is therefore not used as a decision threshold, but rather as contextual evidence that complements saturation and variance by illustrating how mixture complexity is constrained or allowed to evolve across scales.
Taken together, Figures \ref{fig:CrossScalePersistenceMetrics} and \ref{fig:SensitivityAnalysisSaturationvsWindowSizeLI-HT} demonstrate that always-on HTs suppress the natural cross-scale evolution of EM statistical structure, either by anchoring complexity (LI-HT) or by introducing stationary emissions that resist temporal averaging at fine scales (RO-HT). HT-free designs, in contrast, exhibit adaptive behavior characterized by low saturation and higher variability across resolutions.

These findings validate the central detection hypothesis of this work: cross-scale persistence—not absolute spectral magnitude or single-scale complexity—is the key assurance signal for identifying always-on parasitic hardware behavior. By jointly analyzing saturation, variance, and median trends, the proposed framework achieves robust, unsupervised, and reference-free discrimination between benign and HT-inserted designs, even across different classes of always-on HTs.

While the proposed framework demonstrates robust reference-free detection of always-on HTs through cross-scale persistence analysis, two limitations merit discussion. First, the approach relies on EM side-channel measurements, which are inherently sensitive to probe placement, coupling, and environmental noise; although the proposed metrics emphasize statistical persistence rather than absolute magnitude, extreme measurement variability may still impact detection sensitivity. Second, the Gaussian mixture modeling step employs a bounded model order to preserve physical interpretability and prevent over-segmentation of non-Gaussian EM features; while this choice is empirically justified and stable across modest variations, adaptive model-capacity selection remains an area for further investigation. Despite these limitations, the observed cross-scale persistence trends remain stable across multiple window sizes, HT classes, and model-capacity bounds, supporting the generality of the proposed assurance signal.

\section{Conclusions and Future Work}
This paper presented a reference-free framework for detecting always-on HTs through cross-scale analysis of electromagnetic (EM) side-channels. Instead of relying on golden references or HT-specific signatures, the proposed approach exploits how the statistical structure of EM emissions evolves across time–frequency resolutions. By modeling stability-map features with bounded Gaussian Mixture Models (GMMs) and analyzing cross-scale trends in saturation, variability, and median mixture complexity, the framework captures persistence patterns that distinguish benign designs from HT-inserted ones.

Experimental results on AES implementations with two always-on HT classes—a LI-HT and a RO-HT—show that HT-free designs exhibit adaptive, scale-dependent behavior, while always-on HTs suppress this evolution in characteristic ways. Sensitivity analysis across multiple model-capacity bounds confirms that these behaviors are robust and not dependent on parameter tuning. Future work will extend this framework to additional workloads and composite HT scenarios, and explore lightweight online implementations to enable continuous, in-field cyber assurance.

\centering
\textbf{\\Acknowledgement\\}
This work is supported by the Office of Naval Research under Grant N000142312131.

\bibliographystyle{IEEEtran}
\bibliography{IEEEabrv,DCAS2026}

@book{Salmani2018,
  author    = {Hassan Salmani},
  title     = {Trusted Digital Circuits: Hardware Trojan Vulnerabilities, Prevention and Detection},
  publisher = {Springer},
  address   = {Cham, Switzerland},
  year      = {2018},
  doi       = {10.1007/978-3-319-79081-4}
}

@ARTICLE{7994702,
  author={He, Jiaji and Zhao, Yiqiang and Guo, Xiaolong and Jin, Yier},
  journal={IEEE Transactions on Very Large Scale Integration (VLSI) Systems}, 
  title={Hardware Trojan Detection Through Chip-Free Electromagnetic Side-Channel Statistical Analysis}, 
  year={2017},
  volume={25},
  number={10},
  pages={2939-2948},
  doi={10.1109/TVLSI.2017.2727985}}

@INPROCEEDINGS{11195049,
  author={John, Ashwin Koshy and Pitta, Sai Tarrun and Dofe, Jaya and Pandey, Jai Gopal},
  booktitle={2025 IEEE Conference on Communications and Network Security (CNS)}, 
  title={Hardware Trojan Detection with Machine Learning and Power Side-Channels: A Post-Deployment Analysis}, 
  year={2025},
  doi={10.1109/CNS66487.2025.11195049}}

@ARTICLE{9534884,
  author={Sun, Shaofei and Zhang, Hongxin and Cui, Xiaotong and Dong, Liang and Fang, Xing},
  journal={IEEE Transactions on Circuits and Systems II: Express Briefs}, 
  title={Electromagnetic Side-Channel Hardware Trojan Detection Based on Transfer Learning}, 
  year={2022},
  volume={69},
  number={3},
  doi={10.1109/TCSII.2021.3110954}}

@article{cta3980,
author = {Tahghigh, Mahsa and Shiri, Nabiollah},
title = {A new ripple carry adder structure based on a swing-boosted full adder for concurrent error correction in low-resolution pipeline analog-to-digital converters},
journal = {International Journal of Circuit Theory and Applications},
volume = {52},
number = {9},
pages = {4741-4754},
doi = {https://doi.org/10.1002/cta.3980},
eprint = {https://onlinelibrary.wiley.com/doi/pdf/10.1002/cta.3980},
year = {2024}
}

@misc{KeysightInspectorSC4,
  author       = {{Keysight Technologies}},
  title        = {{Inspector SC4 Side-Channel Analysis Platform}},
  howpublished = {\url{https://www.keysight.com/us/en/products/network-test/device-vulnerability-analysis/inspector-sc4-side-channel-analysis.html}},
  note         = {Accessed: Jan. 2026},
  year         = {2026}
}

@INPROCEEDINGS{6657085,
  author={Salmani, Hassan and Tehranipoor, Mohammad and Karri, Ramesh},
  booktitle={2013 IEEE 31st International Conference on Computer Design (ICCD)}, 
  title={On design vulnerability analysis and trust benchmarks development}, 
  year={2013},
  volume={},
  number={},
  pages={471-474},
  doi={10.1109/ICCD.2013.6657085}}

@article{Shakya2017,
author = {Shakya, Bicky and He, Tony and Salmani, Hassan and Forte, Domenic and Bhunia, Swarup and Tehranipoor, Mark},
year = {2017},
month = {03},
pages = {},
title = {Benchmarking of Hardware Trojans and Maliciously Affected Circuits},
volume = {1},
journal = {Journal of Hardware and Systems Security},
doi = {10.1007/s41635-017-0001-6}
}

@article{10.1109/TVLSI.2024.3458892,
author = {Lee, Daehyeon and Lee, Junghee and Jung, Younggiu and Kauh, Janghyuk and Song, Taigon},
title = {Robust Hardware Trojan Detection Method by Unsupervised Learning of Electromagnetic Signals},
year = {2024},
issue_date = {Dec. 2024},
publisher = {IEEE Educational Activities Department},
address = {USA},
volume = {32},
number = {12},
issn = {1063-8210},
url = {https://doi.org/10.1109/TVLSI.2024.3458892},
doi = {10.1109/TVLSI.2024.3458892},
journal = {IEEE Trans. Very Large Scale Integr. Syst.},
month = dec,
pages = {2327–2340},
numpages = {14}
}

@book{Bhunia2018HardwareTrojan,
  editor    = {Swarup Bhunia and Mark M. Tehranipoor},
  title     = {The Hardware Trojan War: Attacks, Myths, and Defenses},
  year      = {2018},
  publisher = {Springer},
  address   = {Cham, Switzerland},
  doi       = {10.1007/978-3-319-68511-3},
  isbn      = {978-3-319-68510-6},
  note      = {eBook ISBN: 978-3-319-68511-3}
}

@techreport{DoDI_5200_50,
  author       = {{Emil Michael}},
  title        = {{ASSURED ACCESS TO TRUSTED MICROELECTRONICS }},
  institution  = {{Office of the Under Secretary of Defense for Research and Engineering}},
  number       = {},
  address      = {},
  year         = {2025}
}

@ARTICLE{10752788,
  author={Sen, Shreyas and Ghosh, Archisman},
  journal={IEEE Solid-State Circuits Magazine}, 
  title={Circuit-Level Techniques for Side-Channel Attack Resilience: A tutorial}, 
  year={2024},
  volume={16},
  number={4},
  pages={96-108},
  doi={10.1109/MSSC.2024.3444740}}

@article{10.1007/s10836-018-5726-9,
author = {Elnaggar, Rana and Chakrabarty, Krishnendu},
title = {Machine Learning for Hardware Security: Opportunities and Risks},
year = {2018},
issue_date = {April     2018},
publisher = {Kluwer Academic Publishers},
address = {USA},
volume = {34},
number = {2},
issn = {0923-8174},
doi = {10.1007/s10836-018-5726-9},
journal = {J. Electron. Test.},
pages = {183–201},
numpages = {19}
}

@article{PrabhakaraRao2025,
  author  = {Prabhakara Rao, Sriram and Balaji, Shree Ram Abayankar and Rahman, Farishta and Suha, Tasneem and Mahfuz, Tanzim and Hossain, Tanvir and Hasan, Mahmudul and Hasan, Md Sakib and Hoque, Tamzidul and Chakraborty, Prabuddha and Ranganathan, Prakash},
  title   = {{Secure and Trustworthy Microelectronics: Vulnerabilities, Solutions, and Trends}},
  journal = {{Journal of Hardware and Systems Security}},
  volume  = {9},
  number  = {3},
  pages   = {163--189},
  year    = {2025},
  month   = dec,
  doi     = {10.1007/s41635-025-00165-x}
}

@book{McLachlanPeel2000,
  author    = {Geoffrey McLachlan and David Peel},
  title     = {Finite Mixture Models},
  publisher = {John Wiley \& Sons, Inc.},
  year      = {2000},
  series    = {Wiley Series in Probability and Statistics},
  isbn      = {9780471006268},
  doi       = {10.1002/0471721182}
}

@article{10.1214/aos/1176344136,
author = {Gideon Schwarz},
title = {{Estimating the Dimension of a Model}},
volume = {6},
journal = {The Annals of Statistics},
number = {2},
publisher = {Institute of Mathematical Statistics},
pages = {461 -- 464},
year = {1978},
doi = {10.1214/aos/1176344136}}

@book{OppenheimSchafer2022,
  author    = {Oppenheim, Alan V. and Schafer, Ronald W.},
  title     = {Discrete-Time Signal Processing},
  edition   = {3rd},
  year      = {2022},
  publisher = {Pearson},
  note      = {Published July 14, 2021},
  isbn      = {9780137549771},
  address   = {Hoboken, NJ}
}

@article{TahghighArXiv2026,
  author  = {Mahsa Tahghigh and Hassan Salmani},
  title   = {Reference-Free Spectral Analysis of {EM} Side-Channels for Always-on Hardware {T}rojan Detection},
  journal = {arXiv preprint arXiv:2601.20163},
  year    = {2026},
  url     = {https://arxiv.org/abs/2601.20163}
}

@INPROCEEDINGS{11014346,
  author={Elahi, Mehdi and Elshamy, Mohamed R. and Badawy, Abdel-Hameed and Fazeli, Mahdi and Patooghy, Ahmad},
  booktitle={2025 26th International Symposium on Quality Electronic Design (ISQED)}, 
  title={Matter: Multi-Stage Adaptive Thermal Trojan for Efficiency \& Resilience Degradation}, 
  year={2025},
  volume={},
  number={},
  pages={1-8},
  doi={10.1109/ISQED65160.2025.11014346}}

@inproceedings{tahghigh2024gmm,
  author  = {M. Tahghigh and H. Salmani},
  title   = {Detecting Hardware Trojans in Manufactured Chips Without Reference: A {GMM}-Based Approach},
  booktitle = {IEEE/ACM International Conference on Computer-Aided Design (ICCAD '24)},
  address = {New York, NY, USA},
  year    = {2024},
  doi     = {10.1145/3676536.3689919}
}

@article{TahghighArXiv2026b,
  author  = {Mahsa Tahghigh and Hassan Salmani},
  title   = {Reference-Free EM Validation Flow for Detecting Triggered Hardware Trojans},
  journal = {arXiv preprint arXiv:2602.03666},
  year    = {2026},
  url     = {https://arxiv.org/abs/2602.03666}
  }

@article{10.1145/3802543,
author = {Elahi, Mehdi and R. Elshamy, Mohamed and A. Badawy, Abdel-Hameed and Patooghy, Ahmad},
title = {SentinelEdge: An Attention-Based Defense for Real-Time Mitigation of Adversarial Thermal Manipulations in System-on-Chips},
year = {2026},
publisher = {Association for Computing Machinery},
address = {New York, NY, USA},
issn = {1539-9087},
url = {https://doi.org/10.1145/3802543},
doi = {10.1145/3802543},
journal = {ACM Trans. Embed. Comput. Syst.}
}

\end{document}